\documentclass[aps,prl,twocolumn,showpacs]{revtex4}
\usepackage{amsmath}
\usepackage{graphicx}
\usepackage{fancyhdr}
\usepackage{color}
\begin{document}


\title{Magnetic Domain Wall Pumping by Spin Transfer Torque}


\author{C. T. Boone}
\affiliation{Department of Physics and Astronomy, University of California, Irvine, CA 92697}

\author{I. N. Krivorotov}
\affiliation{Department of Physics and Astronomy, University of California, Irvine, CA 92697}
\begin{abstract}
We show that spin transfer torque from direct spin-polarized current applied parallel to a magnetic domain wall (DW) induces DW motion in a direction independent of the current polarity. This unidirectional response of the DW to spin torque enables DW pumping -- long-range DW displacement driven by alternating current. Our numerical simulations reveal that DW pumping can be resonantly amplified through excitation of internal degrees of freedom of the DW by the current.
\end{abstract}
\maketitle

Dynamics of magnetic domain walls (DW) \cite{schryer, berger1, beach} excited by spin-polarized electrical current are a sensitive probe of interactions between spin currents and magnetization of itinerant ferromagnets \cite{brataas}. Because DW can be readily detected and manipulated in magnetic nanostructures \cite{groll, yamag, saitoh, klaui, hayashi, thomas, pizzini, hayashi2}, current-induced DW motion provides a convenient testing ground for theories of coupled spin-dependent transport and magneto-dynamics \cite{thia2, tatara1, li, barnes}. Understanding current-induced DW motion is also important for applications in nonvolatile magnetic memory \cite{parkin2}. 

Most studies of DW dynamics induced by spin-polarized current have focused on the current-in-plane (CIP) geometry, in which current flows in the plane of a ferromagnetic film or wire, perpendicular to the DW plane \cite{yamag, saitoh, klaui, hayashi, thomas, thia1, thia2, tatara1, li}. These studies established that long-range translational DW motion is experimentally observed at current densities well below the threshold current \cite{thia2, tatara1} for excitation of DW motion by Slonczewski spin torque (ST) \cite{slonc1}, and thus other types of current-induced torque, such as non-adiabatic ST, were invoked to explain the observations \cite{thia2, tatara1, barnes}. Recent theoretical \cite{khval,rebei} and experimental work \cite{ravel,boone3} explored another geometry for current-induced DW motion, in which spin-polarized current is applied parallel to the DW plane. This geometry can be realized in a current-perpendicular-to-plane (CPP) spin valve shown in Fig. 1(a), where spin-polarized current from a uniformly magnetized fixed layer induces DW motion in the free layer \cite{boone3}.

In this Letter we theoretically examine DW motion in the free layer of a nanowire-shaped CPP spin valve. We find that Slonczewski ST from direct current is sufficient to induce long-range DW displacement, and that non-adiabatic ST \cite{thia2} and field-like ST \cite{brataas} are not necessary for translational DW motion in the CPP geometry. Our analytical and numerical calculations show that ST in the CPP geometry induces DW motion in a direction independent of the current polarity. This type of DW motion is in sharp contrast to current-induced DW dynamics in the CIP geometry, in which the DW velocity changes sign upon reversal of the current flow direction. As a result, alternating current (ac) in the CPP geometry induces unidirectional DW displacement. This DW pumping by ac ST originates from a non-trivial dependence of ST on the angle between the current polarization and the DW magnetization. Calculations of the DW velocity as a function of the ac current frequency, $f$, reveal that DW pumping can be resonantly enhanced via excitation of the DW internal degrees of freedom by the current.

\begin{figure}
\centering
\includegraphics[width=\columnwidth]{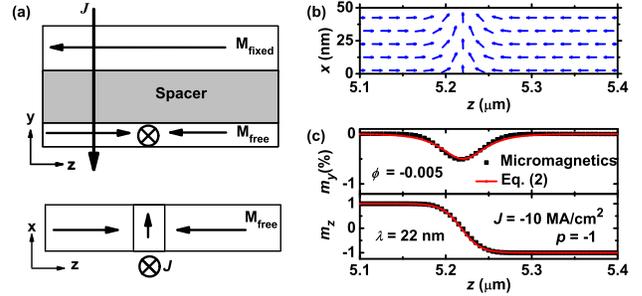}
\caption{(color online) (a) Schematic of a CPP nanowire spin valve with a transverse DW in the free layer.  (b) Magnetization profile of the transverse DW in a 50 nm wide, 5 nm thick Py nanowire given by micromagnetic simulations for the applied current density $J=$-10 MA/cm$^2$. (c) Micromagnetic magnetization profile of the DW ($m_{y}(z)$ and $m_{z}(z)$) at the DW center ($x$ = 25 nm) (squares) and Eq.(\ref{dwshape}) fit with $\lambda$ as the fitting parameter (lines).}
\end{figure}

Motion of magnetization of the free layer in the CPP spin valve shown in Fig. 1(a) is described by the Landau-Lifshitz-Gilbert (LLG) equation with a Slonczewski spin torque term \cite{slonc1, berger, slonc2}:
\begin{equation}
\frac{d\vec{m}}{dt} = -\gamma \vec{m}\times \left(\vec{H}_{e} - \frac{\alpha}{\gamma} \frac{d\vec{m}}{dt} + a_{J} \tau(\Lambda,\vec{m}) (\vec{m}\times \vec{p})\\ \right)
\label{LLGS}
\end{equation}
where $\vec{m}$ is a unit vector in the direction of the free layer magnetization, $\vec{p}$ is a unit vector in the direction of the fixed layer magnetization, $\gamma$ is the gyromagnetic ratio, $\vec{H}_{e}$ is the effective field consisting of Zeeman, anisotropy and exchange terms, $\alpha$ is the Gilbert damping constant, $a_{J} = \frac{\hbar J P}{2deM_s}$ is spin torque parameter, $J$ is the current density, $P$ is the current polarization, $d$ is the free layer thickness, $M_s$ is the free layer saturation magnetization, $e$ is electron charge, and $\tau(\Lambda,\vec{m})\equiv\frac{2\Lambda^2}{(\Lambda^2+1) + (\Lambda^2-1)\vec{m}\cdot\vec{p}}$ describes the dependence of ST on the angle between $\vec{m}$ and $\vec{p}$ \cite{slonc2}. In Eq.(\ref{LLGS}), we omit the field-like ST term \cite{brataas} because it is small in metallic spin valves \cite{brataas,sankey2}.
 
For a thin free layer, the DW is transverse as shown in Fig. 1(b) \cite{beach, nakatani}. The spatial profile of the magnetization of such a DW in the coordinate system of Fig. 1(a) is described by \cite{schryer, thia2, tatara1}:
\begin{subequations}
\begin{eqnarray}
m_x = \sin(\theta) \cos(\phi) = \mbox{sech}\left[\frac{z_0-z}{\lambda}\right] \cos(\phi)\\
m_y = \sin(\theta) \sin(\phi) = \mbox{sech}\left[\frac{z_0-z}{\lambda}\right] \sin(\phi) \\
m_z = \cos(\theta) = \tanh\left[\frac{z_0-z}{\lambda}\right]
\end{eqnarray}
\label{dwshape}
\end{subequations} where $z_0$ is the DW center position, $\phi$ is the out-of-plane angle of magnetization in the DW center and $\lambda$ is the DW width. Figures 1(b) and 1(c) show the DW shape during current-induced motion in a 5 nm thick, 50 nm wide permalloy (Py) nanowire ($M_s$ = 800 emu/cm$^3$, $\alpha=0.01$ \cite{guan} and exchange constant $A= 10^{-6}$ erg/cm) calculated using OOMMF micromagnetic code \cite{donahue}. The DW shape obtained from the micromagnetic simulations is in good agreement with Eq.(\ref{dwshape}). The best fit of  Eq.(\ref{dwshape}) to the micromagnetic data shown in Fig. 1(c) yields $\lambda=$ 22 nm \cite{nakatani}.
 
Substituting Eq.(\ref{dwshape}) into Eq.(\ref{LLGS}), making the rigid domain wall approximation ($\lambda=const=$ 22 nm) and integrating over $z$, we obtain equations of motion for two DW collective coordinates: the DW center position $z_0$ and the DW angle $\phi$ \cite{schryer, tatara1, khval}:
\begin{subequations}
\begin{eqnarray}
\frac{1}{\lambda}\frac{dz_0}{dt}-\alpha\frac{d\phi}{dt} - T= \frac{\gamma}{M_s}(K_y-K_x) \sin(2\phi)\\
-\frac{\alpha}{\lambda} \frac{dz_0}{dt} - \frac{d\phi}{dt} - F = 0
\end{eqnarray}
\label{DWeqmot}
\end{subequations}
where $T$ and $F$ will be called the torque and force drive terms and $K_x$ and $K_y$ are the nanowire shape anisotropy energy constants ($K_x= 0.12\times 2\pi M_s^2$, $K_y= 0.88\times 2\pi M_s^2$) \cite{aharoni}.
For current polarized along the $z$-axis, the expressions for $T$ and $F$:
\begin{subequations}
\begin{eqnarray}
T = p\gamma a_{J} \int^{\infty}_{-\infty}{\tau(\Lambda,m_z)(1-m_z^2)dz}
\label{TQint}
\\
F = p\frac{1}{\pi \cos(\phi)}\gamma a_{J} \int^{\infty}_{-\infty}{\tau(\Lambda,m_z)(m_zm_y)dz}
\label{FcTqintegrals}
\end{eqnarray}
\end{subequations}

can be integrated analytically to give

\begin{subequations}
\begin{eqnarray}
T = p\gamma a_{J}\frac{2 \Lambda^2 \ln[\Lambda]}{\Lambda^2-1}  \equiv p\gamma a_{J} \zeta_1(\Lambda)
\label{Tq}
\\
F = -\gamma a_{J}\frac{\Lambda(\Lambda-1)}{\Lambda + 1}\tan(\phi)\equiv -\gamma a_{J}\zeta_2(\Lambda)\tan(\phi)
\label{Fc}
\end{eqnarray}
\end{subequations}
where $p=\vec{p}\cdot\hat{z}=\pm1$ is the fixed layer magnetization direction.

For angular-symmetric ST ($\Lambda=1$), $F$=0 and Eq. (\ref{DWeqmot}) reduce to the DW equations of motion derived in Ref. \cite{khval}. In this case, ST from direct current gives rise to a small reversible DW displacement and does not induce long-range translational DW motion \cite{khval}. Angular asymmetry of spin torque ($\Lambda>1$) leads to a non-zero force on the DW, which gives rise to a non-zero terminal DW velocity in response to direct current.  The specific form of the ST angular dependence may be different for different spin valve systems \cite{xiao2} but the force term, $F$, is generally non-zero for an angular-asymmetric ST. Figure 2 shows that both the torque $\zeta_{1}(\Lambda)\sim T$ and the force $\zeta_{2}(\Lambda)\sim F$ terms increase with increasing ST angular asymmetry $\Lambda$.

\begin{figure}
\centering
\includegraphics[width=\columnwidth]{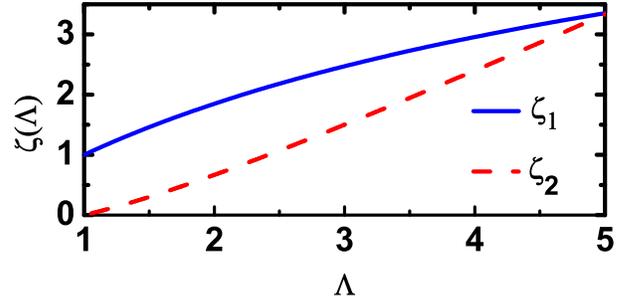}
\caption{(color online) Dependences of the torque $\zeta_1\sim T$ (solid line) and force $\zeta_2\sim F$ (dashed line) terms on ST angular asymmetry parameter $\Lambda$.}
\end{figure}

An analytical expression for the terminal DW velocity under the action of direct current can be derived from Eq. (\ref{DWeqmot}) in the approximation of small DW angle ($\phi<<1$):
\begin{equation}
\frac{dz_0}{dt} = -p\lambda \frac{\gamma M_s \zeta_1(\Lambda)\zeta_2(\Lambda)}{2\alpha (K_y-K_x)} a_{J}^2.
\label{DWvel}
\end{equation}
The quadratic dependence of the DW terminal velocity on current in Eq.(\ref{DWvel}) directly follows from Eq.(\ref{Fc}) stating that the force term $F$ is proportional to the product of $a_J$ and $\tan(\phi)$. Because $\phi\sim a_J$ changes sign upon reversal of the current polarity, $\dot{z_0}\sim F\sim a_J^2$ follows.

To understand the DW dynamics beyond the small-angle approximation, we numerically solve Eq.(\ref{LLGS}) using OOMMF \cite{donahue}. We make micromagnetic simulations of the DW dynamics in a 50 nm wide, 10 $\mu$m long, 5 nm thick Py nanowire for the current polarization $P=$0.30. In our simulations, a current step is applied to the DW in equilibrium, and the location of the DW center, $z_{0}$, is recorded as a function of time. Figure 3(a) shows  $z_{0}(t)$ given by micormagnetic simulations and by numerically solving Eq.(\ref{DWeqmot}) for the current densities $J=\pm15$ MA/cm$^2$ and $\lambda = 22$ nm.  Eq.(\ref{DWeqmot}) and the micromagnetic solution predict similar DW dynamics. We note that the sign of the DW terminal velocity is independent of the current polarity in agreement with Eq.(\ref{DWvel}), and that for $J>0$ the DW velocity changes sign approximately 1.5 ns after application of the current step. 
\begin{figure}[htb]
\includegraphics[width=\columnwidth]{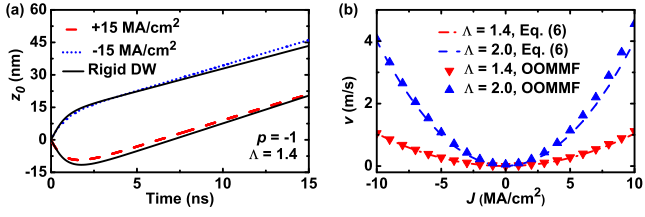}
\caption{(color online) (a) DW position versus time calculated micromagnetically (dotted and dashed lines) and in the rigid DW approximation by solving Eq.(\ref{DWeqmot}) (solid lines) for direct current density $\pm 15$ MA/cm$^2$.  (b) DW terminal velocity as a function of direct current for two values of the ST asymmetry parameter $\Lambda$. Triangles represent micromagnetic results whereas lines are given by Eq.(\ref{DWvel}).}
\end{figure}

Figure 3(b) displays the DW terminal velocity as a function of current for two values of the ST angular asymmetry parameter $\Lambda$. Since the DW terminal velocity is approximately quadratic in the current density, the DW moves in the direction set by the pinned layer magnetization vector, not by the current flow direction. This unusual unidirectional DW dynamics in the CPP geometry enables DW pumping -- persistent translational DW motion in response to alternating current. Figure 4(a) illustrates this translational DW motion for three frequencies of the alternating drive current with an amplitude of 20 MA/cm$^2$.  

\begin{figure}[htb]
\includegraphics[width=\columnwidth]{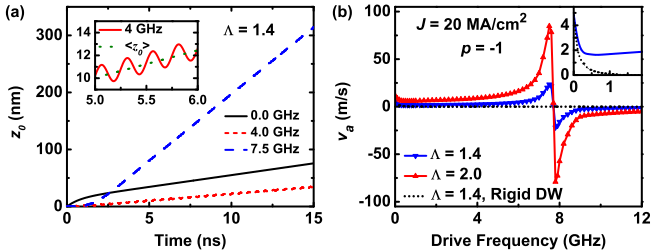}
\caption{(color online) (a) Micromagnetic simulation results for position of the DW center versus time for three frequencies of applied alternating current with an amplitude of 20 MA/cm$^2$. Inset is a blow-up of a 1 ns time interval showing that the DW motion is a combination of oscillatory and translational dynamics. (b) Terminal velocity of ac-driven translational DW motion, $v_a$, as a function of the drive frequency.  The behavior is resonant with the resonance frequency of 7.7 GHz. Inset shows the low-frequency behavior for $\Lambda$=1.4 in greater detail.}
\end{figure}

Figure 4(b) shows the DW terminal velocity, $v_{a}$, averaged over one period of the drive current versus the drive frequency calculated (i) from Eq.(\ref{DWeqmot}) and (ii) from micromagnetic simulations. In the low frequency regime ($f<$ 0.1 GHz), both approaches give similar results -- $v_{a}$ decreases with increasing frequency. This is expected because at short time scales, the instantaneous DW velocity changes its sign upon reversal of the current polarity as illustrated in Fig. 3(a), and thus the net DW displacement is expected to be smaller for ac-driven motion than for dc-driven motion. However, at higher drive frequencies, micromagnetic simulations show a resonant increase in $v_{a}$ (with a resonance frequency of 7.7 GHz), in sharp contrast to the predictions of Eq.(\ref{DWeqmot}). This indicates that the rigid DW approximation assuming the DW shape given by Eq.(\ref{dwshape}) breaks down for high frequencies of the ac drive. 

We analyze the high-frequency resonant motion of DW driven by alternating current in the DW center-of-mass reference frame (COMF), defined as a the reference frame where $z_0 =const\equiv$0. Figure 5(a) shows the out-of-plane component of magnetization, $m_y$, in the middle of the wire ($x$ = 25 nm) as a function of time and distance along the wire in the DW COMF.  The magnetization driven by alternating current at a frequency $f=$7.5 GHz with amplitude 20 MA/cm$^2$ exhibits oscillatory behavior, and Fourier analysis shows that the resulting dynamics consist of motions at three frequencies.  Figure 5(b) shows $\tilde m_y(z,f)$ -- the Fourier transform of $m_y(z,t)$ in the COMF at three positions along the wire length ($z$).  The exact center of the DW ($z=0$, $x=25$nm) is dominated by a 0 GHz mode, the region 50 nm from the center is dominated by oscillations at $7.5$ GHz ($f$), and the region far from the DW ($z=500$ nm) is dominated by $15$ GHz ($2f$) oscillations.

\begin{figure}[htb]
\includegraphics[width=\columnwidth]{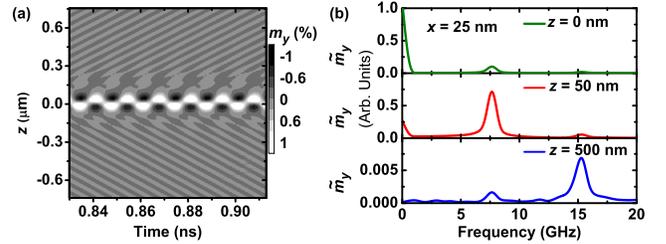}
\caption{(color online) (a) Out-of-plane component of magnetization, $m_y(z,t)$, at $x=25$ nm plotted in the DW COMF reveals spatially antisymmetric oscillations near the DW center, and propagating spin waves far from the DW center for DW driven by 20 MA/cm$^2$, 7.5 GHz alternating current. (b) Fourier transform of $m_y(z,t)$, $\tilde m_y(z,f)$, at three different locations along the wire length ($z=$ 0, 50 and 500 nm). The DW center ($z=0$) is dominated by 0 GHz translational DW mode. An antisymmetric localized DW mode with a frequency of 7.5 GHz is dominant at $z$= 50 nm. Dynamics far away form the DW center are characterized by propagating spin waves with a frequency of 15 GHz and a wavelength of 65 nm.}
\end{figure}

The DW velocity resonances with frequencies near 0 GHz and 7.7 GHz are due to excitation of the lowest (symmetric) and second-lowest (antisymmetric) bound spin wave modes, which are localized in the potential well formed by the DW \cite{herms,han}. The 0 GHz mode is the usual translational DW mode with $\dot{\phi}\approx0$ \cite{winter} that also exists in the rigid DW approximation. The 7.7 GHz mode is a spatially antisymmetric localized mode \cite{han,winter}. Excitations at 15 GHz are propagating spin waves emitted by the oscillating DW at twice the drive frequency \cite{herms}. Fig. 6 illustrates the spatial profiles of the symmetric ($\vec{m}_{s}$) and antisymmetric ($\vec{m}_{a}$) modes. 

For a drive frequency close to 7.7 GHz, the DW exhibits enhanced translational velocity, $v_a$ due to excitation of the antisymmetric mode by the alternating current. Indeed, Eq.(\ref{FcTqintegrals}) shows that the force term, $F$, that gives rise to long-range translational DW motion becomes large for the antisymmetric mode because $m_zm_y$ becomes approximately even in $z$, making the integral over $z$ in Eq.(\ref{FcTqintegrals}) large. The resonant enhancement of $v_a$ takes place because the drive frequency must be close to the resonance frequency for the amplitude of the antisymmetric mode (and thus the force term $F$) to be large. The terminal DW velocity changes sign as the drive frequency is swept through the frequency of the antisymmetric mode (7.7 GHz) because the phase between the ac ST drive, $a_J$, and the out-of-plane component of magnetization, $m_y$, of the antisymmetric mode changes by 180$^\circ$ when the drive frequency is swept through the resonance frequency. The 180$^\circ$ phase shift leads to the sign change of the force term, $F$, in Eq.(\ref{FcTqintegrals}) and thus to the sign change of the DW terminal velocity, $v_a$.

\begin{figure}[htb]
\includegraphics[width=\columnwidth]{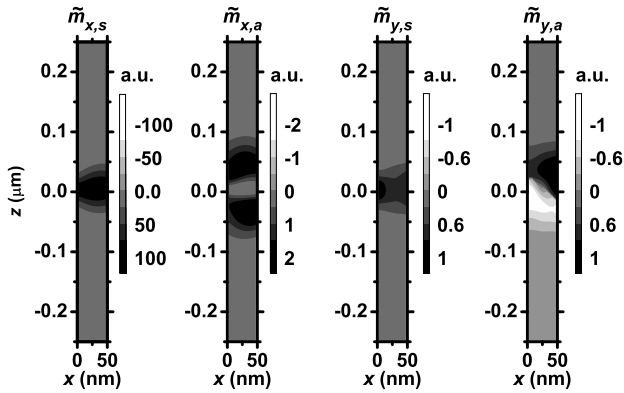}
\caption{Numerically determined spatial profiles of the symmetric ($\vec{m}_{s}$, 0 GHz) and antisymmetric ($\vec{m}_{a}$, 7.7 GHz) localized DW modes.}
\end{figure}

In conclusion, we have shown that spin transfer torque from direct current in the CPP geometry gives rise to persistent translational DW motion. In this geometry, the terminal DW velocity is independent of the current polarity, which gives rise to DW pumping -- long-range DW displacement driven by alternating spin torque. Our numerical simulations reveal that DW pumping is resonantly amplified when frequency of the ac spin torque drive is close to the frequency of an antisymmetric localized DW mode. This work was supported by NSF Grants DMR-0748810 and ECCS-0701458 and by the Nanoelectronics Research Initiative through the Western Institute of Nanoelectronics.

\end{document}